\documentclass[conference]{IEEEtran}
\usepackage{fancyhdr}
\usepackage[utf8]{inputenc}
\usepackage{cite}
\usepackage{tikz}
\usetikzlibrary{3d}
\usepackage{pgfplots}
\usepackage{listings}
\usepackage[caption=false]{subfig}
\usepackage{amsmath,amssymb,amsfonts}
\usepackage{algorithm}
\usepackage[]{algpseudocode}
\usepackage{graphicx}
\usepackage{pdflscape}
\usepackage{color}
\usepackage{xcolor}
\usepackage[T1]{fontenc}
\usepackage[scaled]{beramono}
\usepackage[driver=pdftex, margin=2.54cm]{geometry}
\usepackage{textcomp}
\usepackage{multirow}
\usepackage{comment}
\def\BibTeX{{\rm B\kern-.05em{\sc i\kern-.025em b}\kern-.08em
    T\kern-.1667em\lower.7ex\hbox{E}\kern-.125emX}}
\pgfplotsset{compat=1.16}

\makeatletter
\lst@InstallKeywords k{attributes}{attributestyle}\slshape{attributestyle}{}ld
\makeatother

\definecolor{bluekeywords}{rgb}{0.2,0.2,0.7}
\definecolor{graycomments}{rgb}{0.3,0.3,0.3}
\definecolor{redstrings}{rgb}{0.74,0.08,0.08}
\definecolor{background}{rgb}{0.95,0.95,0.95}
\definecolor{types}{rgb}{0.17,0.57,0.68}

\begin{document}

\title{Fast Stencil-Code Computation on a Wafer-Scale Processor}



\author{
\IEEEauthorblockN{Kamil Rocki\IEEEauthorrefmark{1}, 
Dirk Van Essendelft\IEEEauthorrefmark{2}, 
Ilya Sharapov\IEEEauthorrefmark{1}, 
Robert Schreiber\IEEEauthorrefmark{1}, 
Michael Morrison\IEEEauthorrefmark{1}, \\
Vladimir Kibardin\IEEEauthorrefmark{1}, 
Andrey Portnoy\IEEEauthorrefmark{1}, 
Jean Francois Dietiker\IEEEauthorrefmark{2}\IEEEauthorrefmark{3}, 
Madhava Syamlal\IEEEauthorrefmark{2} 
and Michael James\IEEEauthorrefmark{1}} \\
\IEEEauthorblockA{
\IEEEauthorrefmark{1}
Cerebras Systems Inc., 
Los Altos, California, USA\\
Email: \{kamil,michael\}@cerebras.net} \\
\IEEEauthorblockA{
\IEEEauthorrefmark{2}
National Energy Technology Laboratory, 
Morgantown, West Virginia, USA\\
Email: dirk.vanessendelft@netl.doe.gov} \\
\IEEEauthorblockA{
\IEEEauthorrefmark{3}
Leidos Research Support Team, 
Pittsburgh, Pennsylvania, USA\\
Email: jean.dietiker@netl.doe.gov}

}

\vspace{0.5cm}

\maketitle
\thispagestyle{fancy}
\lhead{}
\rhead{}
\chead{}
\lfoot{\footnotesize{
SC20, November 9-19, 2020, Is Everywhere We Are
\newline 978-1-7281-9998-6/20/\$31.00 \copyright 2020 IEEE}}
\rfoot{}
\cfoot{}
\renewcommand{\headrulewidth}{0pt}
\renewcommand{\footrulewidth}{0pt}
\begin{abstract}
The performance of CPU-based and GPU-based systems is often low for PDE codes, where large, sparse, and often structured systems of linear equations must be solved. Iterative solvers 
are limited by data movement, both between caches and memory and between nodes. Here we describe the solution of such systems of equations
on the Cerebras Systems CS-1, a wafer-scale processor that has the memory bandwidth and communication latency to perform well. We achieve 0.86 PFLOPS on a single wafer-scale system for the solution by BiCGStab of a linear system arising from a 7-point finite difference stencil on a $600  \times 595 \times 1536$ mesh, achieving about one third of the machine's peak performance.  We explain the system, its architecture and programming, and its performance on this problem and related problems.  We discuss issues of memory capacity and floating point precision.  We outline plans to extend this work towards full applications.
\end{abstract}


\section{Introduction}
The need for high memory bandwidth is captured by a problem's {\em arithmetic intensity}, the number of operations performed on each datum loaded from memory.  
Solvers of partial differential equations by finite difference, element, or volume methods have low intensity, as they make repeated sweeps over meshes.
Performance for them on CPU or GPU based systems suffers due to insufficient bandwidths. For example, on the high-performance conjugate gradient (HPCG) benchmark, the top 20 performing supercomputers achieve only 0.5\% - 3.1\% of their peak floating point performance~\cite{hpcg}. For HPCG as for many HPC kernels and real applications, limited memory bandwidth and high communication latency are primary performance limiters. 

\begin{figure}[htbp]
\centerline{\includegraphics[width=0.47\textwidth]{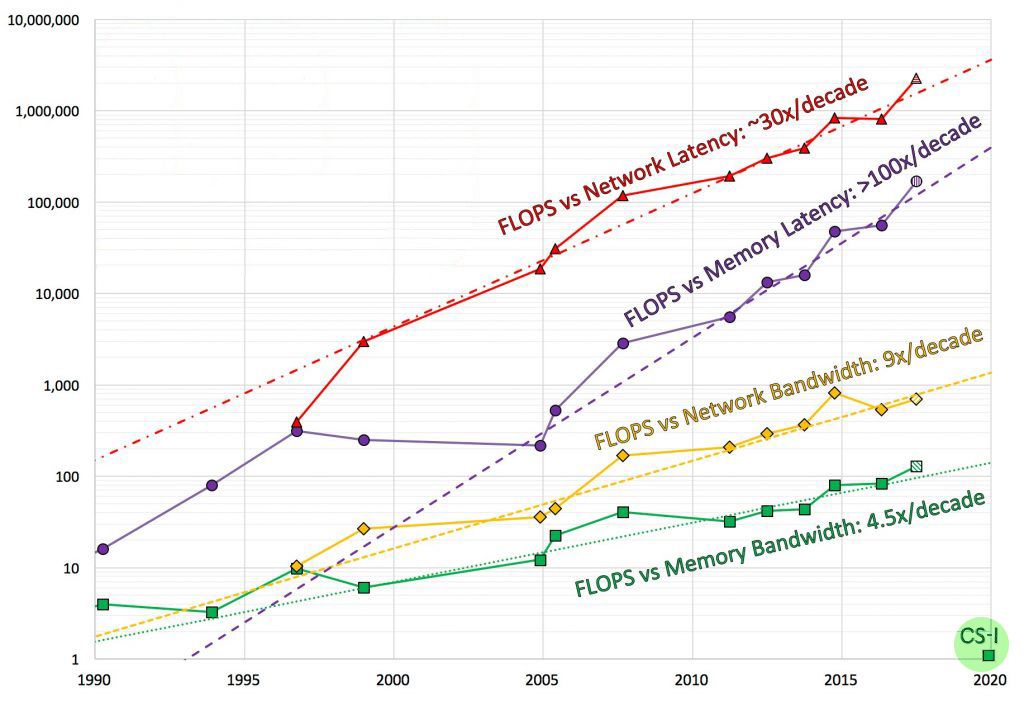}}
\caption{The growing gulf in flops per word (memory, interconnect) of conventional CPUs and clusters, and the impact of wafer-scale integration. (Figure courtesy of John McCalpin~\cite{mccalpin16}, used with permission.  CS-1 data point added by us.) }
\label{fig:mccalpin}
\end{figure}

HPC memory and communication systems struggle to keep up with processing performance. In 2016 the flops to words ratios for both memory and interconnect bandwidth were in the hundreds, and the flops needed to cover the memory or network latencies were in the 10,000 to 100,000 range, with the trend going higher; see Figure~\ref{fig:mccalpin}.

The recent introduction of wafer-scale processors promises to change this situation. The Cerebras Systems CS-1, a system whose compute and memory resources are all fabricated in a single $462~\textrm{cm}^2$ silicon wafer, can move three bytes to and from memory for every flop.  This is achieved in a highly parallel, distributed memory architecture in which all memory is on the same silicon wafer as the processing, providing orders of magnitude more memory bandwidth, single cycle memory latency, and lower energy cost for memory access.   There are 18 GB of on-wafer memory. The memory on the wafer is fast, static random-access memory (SRAM). Processing elements are interconnected by an on-the-wafer network with injection bandwidth one fourth of the peak floating point compute bandwidth and with nanosecond per hop message latencies.  Thus, on the \mbox{CS-1}, memory bandwidth matches the peak compute rate, and communication bandwidth is only slightly lower.
The CS-1, as shown in Figure~\ref{fig:mccalpin}, sits at the desirable bottom on the flops per access scale, a region that becomes ever farther away as single chip systems are pushed towards greater on chip capability with the hard limits of off-chip communication.  

Here we demonstrate the potential for wafer-scale systems to achieve breakthrough performance on regular mesh finite difference (stencil) problems that can fit in the on-wafer memory, by giving some examples of the implementation and performance of model problems on the CS-1.
We have implemented a BiCGStab solver for a linear system arising from the 7-point discretization of a PDE on a 3D mesh.  
The solver achieves performance of 0.86 PFLOPS in mixed precision floating point that uses 16-bit for all arithmetic except the inner products and a mixed precision inner product with 16-bit multiply and 32-bit add.   
The achieved performance per Watt (at 20 kW) and for the size of the machine (1/3 rack) are beyond what has been reported for conventional machines on comparable problems. 

We describe the system, the processing core architecture, the programming model, and our implementation of BiCGStab in that model.  
We discuss the implementation of the global reductions required for, among other things, the inner products common to all Krylov subspace linear solvers that are often a bottleneck in large-scale message-passing implementations.  Because on-wafer communication has low latency, our AllReduce (to use the MPI term) for scalars takes under 1.5 microseconds for a system of about 380,000 independent, interconnected processors.
We then present and validate a simple performance model, and use it to predict the effect of changing mesh size and shape and of an implementation for a problem arising from a large two-dimensional mesh. 

Then we talk about the limitations of this experiment and potential extensions to more complex and realistic applications. 
We sketch a proposed implementation of the NETL\footnote{National Energy Technology Laboratory} code, MFIX, for modeling combustion or chemical reactions of solid particles (e.g., fuels, sorbents) transported in a fluid.  The performance projections for MFIX indicate that real-time, highly resolved simulation will be possible, opening many opportunities for practical applications that demand both high fidelity and speed.  

We then consider issues of floating point arithmetic precision and show the accuracy achieved with mixed (16/32) and higher precision for this problem.  

Finally, we discuss the limitation of the single-wafer solution (a limited amount of memory), pointing out situations where the speed is critical and the size is adequate, and indicating the evolution of the technology towards greater capacity.
\section{The CS-1 Wafer Scale Engine}

\begin{figure*}[hbt!]\centering
\includegraphics[width=0.95\textwidth]{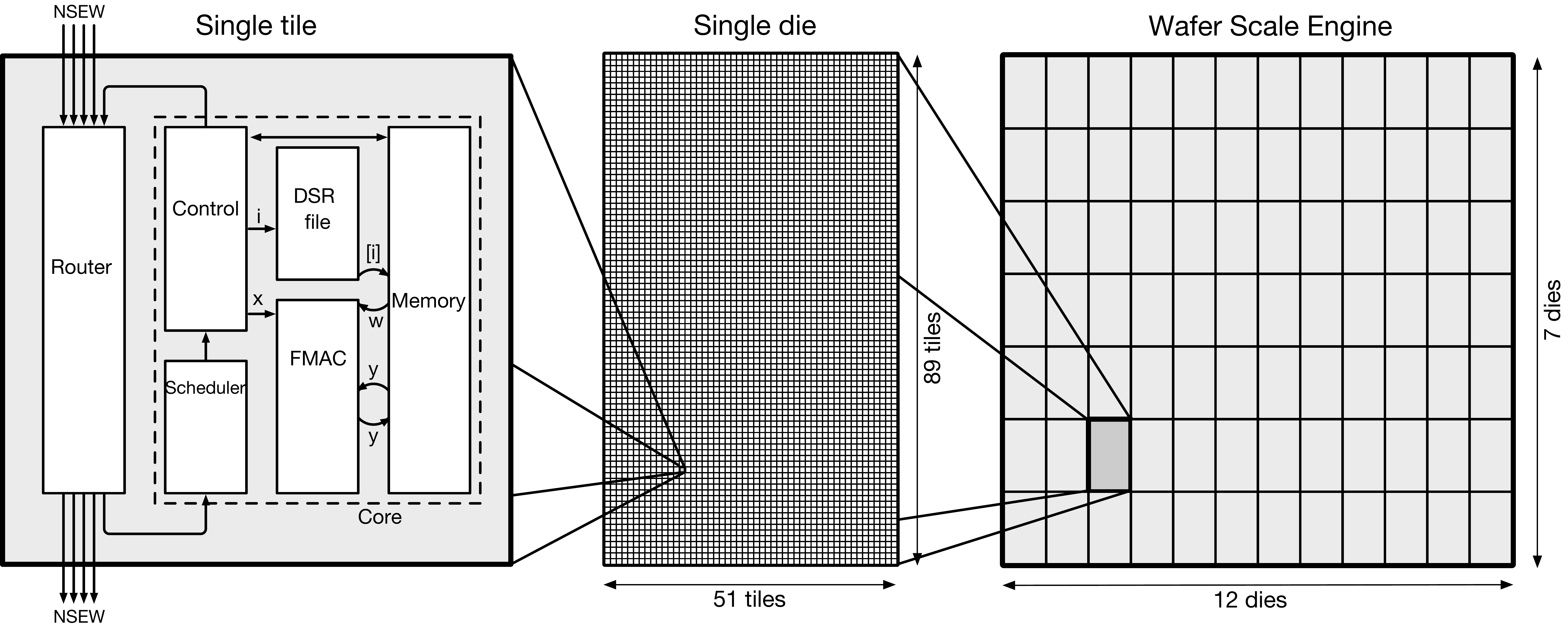}
\caption{CS-1 Wafer Scale Engine (WSE). A single wafer (rightmost) contains one CS-1 processor. Each processor is a collection of dies arranged in a 2D fashion (middle). Dies are then further subdivided into a grid of tiles. One die hosts thousands of computational cores, memory and routers (leftmost). There is no logical discontinuity between adjacent dies and there is no additional bandwidth penalty for crossing the die-die barrier. In total, there are 1.2 trillion transistors in an area of 462.25 $cm^{2}$.}
\label{fig:cs1:wafer}
\end{figure*}

Cerebras Systems, formed in 2016, has designed and brought to market the industry's first wafer-scale system~\cite{sean-hotchips-task-on-the-web}.  
All compute and memory resources of the \mbox{CS-1} are contained on the Wafer-Scale Engine, a $462~\mathrm{cm}^2$ silicon wafer.  The wafer also contains a powerful communication network; see Figure~\ref{fig:cs1:wafer}.   The system comprises 380,000 processor cores, each with 48 KB of dedicated SRAM memory (for a total of 18GB); up to eight 16-bit floating point operations per cycle; 16 bytes of read and 8 bytes of write bandwidth to the memory per cycle; a 2D mesh interconnection fabric with 16 bytes of injection bandwidth per core per cycle; and a total system power of 20 kW.

\subsection{Architecture}
The CS-1 wafer is an MIMD, distributed-memory machine with a 2D-mesh interconnection fabric.  
The repeated element of the architecture is called a tile.  The tile contains one processor core, its memory, and the router that it connects to. The routers link to the routers of the four neighboring tiles.  Figure~\ref{fig:cs1:wafer} illustrates the layout. 

The wafer contains a $7 \times 12$ array of 84 identical ``die.''  
A die holds thousands of tiles.  Ordinary chips are made by cutting the wafer into individual die; in the WSE, the die are instead connected by extending the interconnect across the "scribe lines", the spaces between die.  

The memory, functional units, and instruction set are designed for high throughput numerical computation.  The roughly 380,000 tiles each have their own fast SRAM memory.   There is no shared memory.  Local memory is 48 KB, which totals 18 GB across the wafer.  The load-to-use latency is one cycle.  

The instruction set supports operations on 16-bit integer, 16-bit (IEEE fp16), and 32-bit (IEEE fp32) floating point types.  
Floating point adds, multiplies, and fused multiply-accumulate (or FMAC, with no rounding of the product prior to the add) can occur in a 4-way SIMD manner for 16-bit operands.  
The instruction set supports SIMD operations across subtensors of four dimensional tensors, making use of tensor address generation hardware to efficiently access tensor data in memory.   These play the role of nested loops and eliminate any loop overhead.
There are enough memory banks to provide the bandwidth needed to fetch eight 16-bit words from memory and store four such words per cycle, enough to support SIMD-4, AXPY operations $\mathbf{y} = \mathbf{y} + a \times \mathbf{x}$, where the operand $a$ is a scalar held in a register and $x$ and $y$ are tensors that stream to and from memory. Such an operation can be launched with a single instruction.  The tensor operands can have more than four elements, so the instruction executes for multiple cycles.

In mixed precision with multiplications in fp16 and additions performed in fp32, the throughput is two FMACs per core per cycle.  Purely 32-bit floating point computations run one FMAC per core per cycle.  The theoretical peak performance of the system varies depending on the number of cores configured on the wafer, clock rate and power settings.

The core connects to a local router that has five bidirectional links, one to each of its four nearest neighbors and one to its own core.  The router can move data into and out of these five links, in parallel, on every cycle. Even with scalar granularity, communication is efficient.  
The router has hardware queues for its connection to the core and for each of a set of virtual channels, avoiding deadlock. 
Communication between potentially distant processors occurs along predetermined routes.  
Routing is configured offline, as part of compilation; data travel along virtual channels that can be programmatically reconfigured at run time.  The fanout of data to multiple destinations is done through the routing; the router can forward an input word to any subset of its five output ports.
There is no runtime software involved with communication.  Arriving data are deposited by the hardware directly into memory or registers or routed to functional units as specified by the program.

An instruction with tensor operands can run synchronously or, at the discretion of the programmer, as a background thread that shares the datapath with other threads including the main one.  A background thread runs a single tensor operation, as a single asynchronously running instruction.  There is no context switch overhead. The registers and memory used by an asynchronous thread are those assigned by the programmer or compiler in the instruction, and these may not be overwritten until the thread terminates.  Subsequent computation can be delayed until the thread terminates.  The core supports nine concurrent threads of execution. 

A stream of data to or from the fabric may be used as an input to a tensor operation, or as the destination for one.  The hardware directly implements scheduling activities that would normally be performed by an operating system. This allows compact and efficient software implementations. For example, one core can be sending data from its local memory to another core; simultaneously it can receive data from another core while adding it to values stored in its local memory. All of this is accomplished using only two machine instructions that run as independent threads.


Code consists of tasks that react to events.  Tasks are triggered by other tasks, or by arriving data words.  The channel of the arriving word determines the code that is triggered.   There is little delay between the completion of a task and the start of a subsequent task, as this is handled in hardware. 
Together with the SIMD operations of the instruction set, this efficient scalar-data-triggered method of operation allows us to implement distributed linear algebra kernels, such as those used in machine learning and the iterative linear solver discussed here, with minimal performance impact from message latency and bandwidth.  Special purpose Data Structure Registers (DSRs) generate tensor access addresses in hardware eliminating overheads of nested loops. 

\section{The BiCGstab Method}

Discretized partial differential equations lead to systems of linear equations 
\begin{equation}
\begin{aligned}
\label{eq:system}
A x = b
\end{aligned}
\end{equation} 
that are commonly solved using Krylov subspace iterative methods such as the conjugate gradient (CG) method.  The Biconjugate Gradient Method\cite{templates} extends CG to nonsymmetric systems. The stabilized version of the method, BiCGStab\cite{vanderVorst1992} (Algorithm~\ref{alg:BiCGStab}), makes it numerically stable (and uses four dot products per iteration instead of two). The kernel operations in the algorithm are sparse matrix - dense vector multiply (SpMV), AXPY ($y = y + ax$ for vectors $x$ and $y$ and scalar $a$), and inner product. In SpMV, each matrix element is involved in only one multiply-add operation; thus the arithmetic intensity is low.

\begin{algorithm}
\caption{Standard BiCGStab}
\label{alg:BiCGStab}
\begin{algorithmic}[1]

\Function{BiCGStab}{$A,b,x_0$}
    \State $r_0 := b, p_0:=r_0$
    \For{i = 0,1,2, ...}
        \State $s_i:=Ap_i$
        \State $\alpha_i := \frac{(r_0,r_i)}{(r_0,s_i)}$
        \State $q_i:=r_i - \alpha_i s_i$
        \State $y_i:=Aq_i$
        \State $\omega_i := \frac{(q_i,y_i)}{(y_i,y_i)}$
        \State $x_i:=x_i + \alpha_i p_i + \omega_i q_i$
        \State $r_{i+1}:=q_i - \omega_i y_i$
        \State $\beta_i := \frac{\alpha_i}{\omega_i} \frac{(r_0,r_{i+1})}{(r_0, r_i)}$
        \State $p_{i+1}:=r_{i+1} + \beta_i (p_i - \omega_i s_i)$
    \EndFor
\EndFunction

\end{algorithmic}
\end{algorithm}

\section{Mapping BiCGstab to the CS-1}

The CS-1 architecture provides a good match to the compute, memory, and communication needs of Krylov subspace methods, especially on a regular mesh.  We map our test case's regular 3D mesh to the 2D machine in a straightforward domain decomposition manner.
Let the mesh be $X \times Y \times Z$. Then map $X$ and $Y$ across the two axes of  
the fabric, with each core handling all of the $Z$ dimension (Figure~\ref{fig:cubes}).  The mesh mapping dictates the mapping of the vectors in the BiCGstab method as each vector element is associated with one meshpoint.  As to the matrix  $A$, we map the needed portion of its nonzero diagonals to each core. $A$ has seven nonzero diagonals; but with diagonal preconditioning the main diagonal is all ones.  Therefore, we only store six other diagonals.   In addition to the elements of $A$, local portions of four vectors must be stored to implement the BiCGstab iteration.  The vectors $q_i$ and $r_{i+1}$ reuse the storage of $s_i$ and $y_i$, leading to a storage requirement per core of $10Z$ words.  Thus, with $Z = 1536$ we are using about 31KB out of 48KB for the matrix and vector data.


\subsubsection{SpMV (3D)}
\begin{figure*}[htbp]
\centerline{\includegraphics[width=0.88\textwidth]{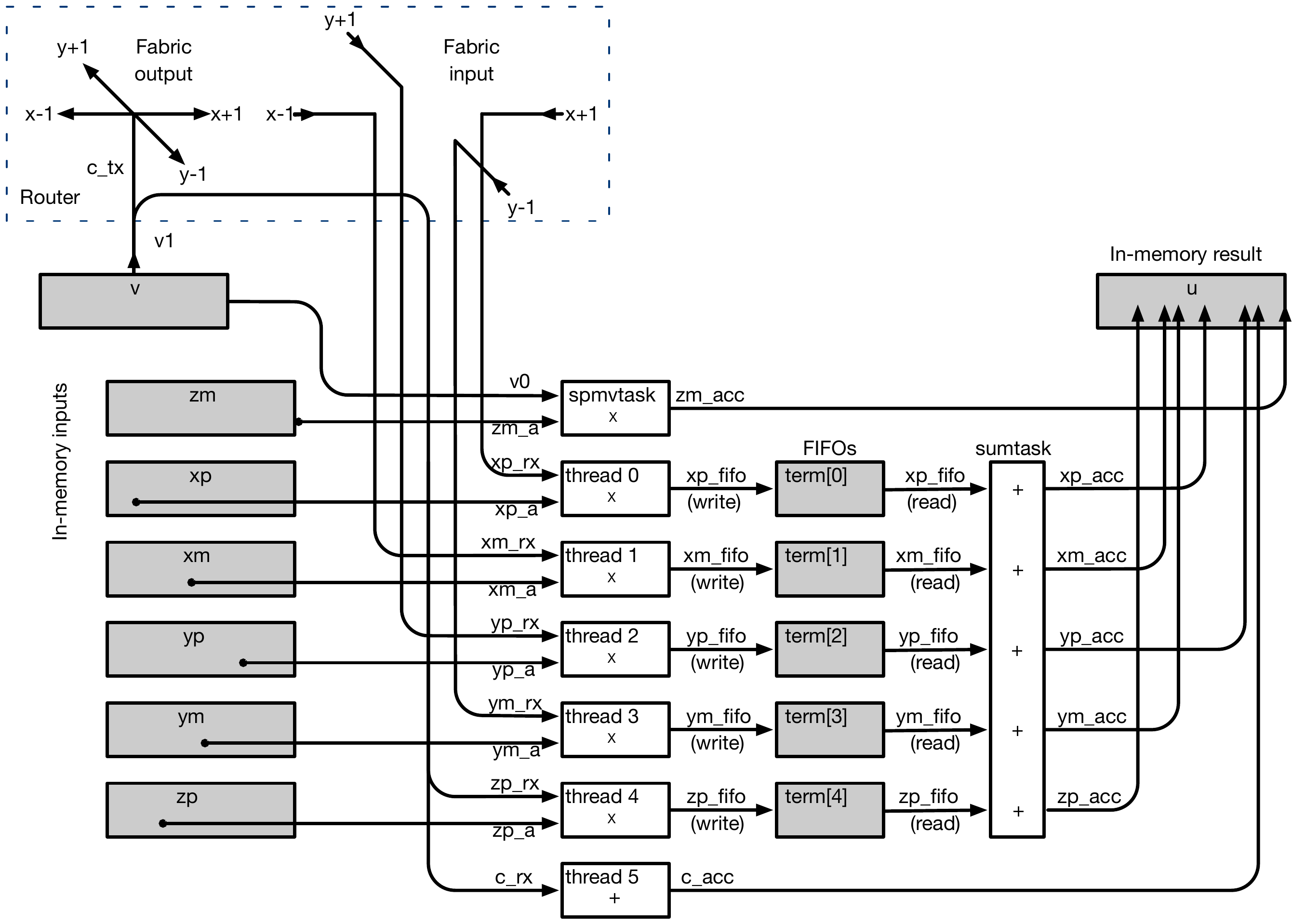}}
\caption{Implementation of SpMV ($u = Av$).  Shaded regions represent memory objects, annotated arrows are tensor descriptors, and white boxes are tasks that perform computations.  The diagram uses the names of objects in the code of Listing 1.}
\label{fig:spmv_flow}
\end{figure*}

The SpMV implementation uses architectural concepts that may be unfamiliar.  To convey how it works we provide a pseudocode listing (Linsting~1) and a coordinated diagram showing the data flow (Figure~\ref{fig:spmv_flow}).    The code allocates memory for vectors, specifies the communication connectivity between tiles, defines tensor descriptors, and specifies tasks and threads that constitute a specification of data flow that is illustrated in the figure.

\begin{figure}[htbp]
\begin{center}
\newcommand{\xangle}{10}
\newcommand{\yangle}{147}
\newcommand{\zangle}{90}

\newcommand{\xlength}{0.94}
\newcommand{\ylength}{0.75}
\newcommand{\zlength}{1}

\newcommand{\h}{-1.5}
\newcommand{\e}{0.4} 
\newcommand{\m}{0.1} 
\newcommand{\opc}{0.5}

\newcommand{\dimension}{2}

\pgfmathsetmacro{\xx}{\xlength*cos(\xangle)}
\pgfmathsetmacro{\xy}{\xlength*sin(\xangle)}
\pgfmathsetmacro{\yx}{\ylength*cos(\yangle)}
\pgfmathsetmacro{\yy}{\ylength*sin(\yangle)}
\pgfmathsetmacro{\zx}{\zlength*cos(\zangle)}
\pgfmathsetmacro{\zy}{\zlength*sin(\zangle)}

\begin{tikzpicture}
[   x={(\xx cm,\xy cm)},
    y={(\yx cm,\yy cm)},
    z={(\zx cm,\zy cm)},
]

\pgfmathsetmacro{\w}{100}
\foreach \a in {0,...,\dimension}{
    \foreach \b in {0,...,\dimension}{
         \draw[canvas is xy plane at z=\a, black!\w][opacity=\opc] (\b,0) -- (\b,\dimension);
         \draw[canvas is xz plane at y=\a, black!\w][opacity=\opc] (0,\b) -- (\dimension,\b);
         \draw[canvas is yz plane at x=\a, black!\w][opacity=\opc] (\b,0) -- (\b,\dimension);
    }
}

\foreach \a in {0,...,\dimension}{
    \foreach \b in {0,...,\dimension}{
         \draw[canvas is xy plane at z=\h, black!\w][opacity=\opc]
               (\a-\e,\b-\e) -- (\a-\e,\b+\e)
               (\a+\e,\b-\e) -- (\a+\e,\b+\e)
               (\a-\e,\b-\e) -- (\a+\e,\b-\e)
               (\a-\e,\b+\e) -- (\a+\e,\b+\e);
    }
    \draw[canvas is xy plane at z=\h, black!\w][opacity=\opc]
        (\a-\e,-1+\e-\m) -- (\a-\e,-1+\e)
        (\a-\e,-1+\e) -- (\a+\e,-1+\e)
        (\a+\e,-1+\e-\m) -- (\a+\e,-1+\e);
    \draw[canvas is xy plane at z=\h, black!\w][opacity=\opc]
        (\a-\e,\dimension+1-\e+\m) -- (\a-\e,\dimension+1-\e)
        (\a-\e,\dimension+1-\e) -- (\a+\e,\dimension+1-\e)
        (\a+\e,\dimension+1-\e+\m) -- (\a+\e,\dimension+1-\e);
    \draw[canvas is xy plane at z=\h, black!\w][opacity=\opc]
        (-1+\e-\m,\a-\e) -- (-1+\e,\a-\e)
        (-1+\e,\a-\e) -- (-1+\e,\a+\e)
        (-1+\e-\m,\a+\e) -- (-1+\e,\a+\e);
    \draw[canvas is xy plane at z=\h, black!\w][opacity=\opc]
        (\dimension+1-\e+\m,\a-\e) -- (\dimension+1-\e,\a-\e)
        (\dimension+1-\e,\a-\e) -- (\dimension+1-\e,\a+\e)
        (\dimension+1-\e+\m,\a+\e) -- (\dimension+1-\e,\a+\e);
}

\draw[canvas is xy plane at z=\h, black!\w][fill=gray!30]  
      (0-\e,1-\e) -- (0-\e,1+\e) -- (0+\e,1+\e) -- (0+\e,1-\e) -- (0-\e,1-\e)
      (2-\e,1-\e) -- (2-\e,1+\e) -- (2+\e,1+\e) -- (2+\e,1-\e) -- (2-\e,1-\e)
      (1-\e,1-\e) -- (1-\e,1+\e) -- (1+\e,1+\e) -- (1+\e,1-\e) -- (1-\e,1-\e)
      (1-\e,0-\e) -- (1-\e,0+\e) -- (1+\e,0+\e) -- (1+\e,0-\e) -- (1-\e,0-\e)
      (1-\e,2-\e) -- (1-\e,2+\e) -- (1+\e,2+\e) -- (1+\e,2-\e) -- (1-\e,2-\e);

\begin{scope}[canvas is yz plane at x=1]
   \draw [line width=0.1mm,->,opacity=\opc] (0,0) -- (0,\h);
   \draw [line width=0.1mm,->,opacity=\opc] (1,0) -- (1,\h);
   \draw [line width=0.1mm,->,opacity=\opc] (2,0) -- (2,\h);
\end{scope}
\begin{scope}[canvas is xz plane at y=1]
   \draw [line width=0.1mm,->,opacity=\opc] (0,0) -- (0,\h);
   \draw [line width=0.1mm,->,opacity=\opc] (2,0) -- (2,\h);
\end{scope}

\foreach \a in {0,...,\dimension}{
    \foreach \b in {0,...,\dimension}{
         \foreach \c in {0,...,\dimension}{
         }
    }
}

\shade[ball color=black!50!white] (1,1,1) circle (.1cm);
\shade[ball color=black!50!white] (0,1,1) circle (.1cm);
\shade[ball color=black!50!white] (2,1,1) circle (.1cm);
\shade[ball color=black!50!white] (1,0,1) circle (.1cm);
\shade[ball color=black!50!white] (1,2,1) circle (.1cm);
\shade[ball color=black!50!white] (1,1,0) circle (.1cm);
\shade[ball color=black!50!white] (1,1,2) circle (.1cm);

\end{tikzpicture}
\end{center}
\caption{Three dimensional problem mapping to two dimensional fabric of processing elements}
\label{fig:cubes}
\end{figure}
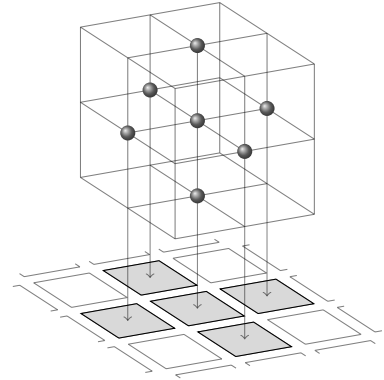

With the iterate vector $v(x,y,z)$ mapped as shown in Figure~\ref{fig:cubes}, each core stores a local iterate vector in an array of $v$ of length $Z$.  The local result of the SpMV is an array $u$ of length $Z$ computed as the sum of seven vectors. Six of these are elementwise products of a vector of matrix elements and a vector of iterate elements. Four of the iterate vectors stream in from the neighboring cores.  Two are shifted-by-one (in the $Z$ direction) copies of the local iterate.   Thus, to perform an SpMV we use eight memory vectors.

The eight memory vectors, shown in gray, are declared with the \texttt{float16} keyword in the code.  
There are the local iterate (\texttt{v}), the result vector (\texttt{u}), and six vectors of matrix elements, denoted by coordinate directions (\texttt{x}, \texttt{y}, \texttt{z}) and \texttt{p} or \texttt{m} to indicate plus or minus, according to their role in the 7-point stencil. In order to make an asynchronous
\newpage
\begin{lstinputlisting}[breaklines=true,caption={SpMV listing}]{spmv.c}
\end{lstinputlisting}

\noindent
connection from multiplication threads to addition threads, the code allocates five in-memory FIFOs to forward the elementwise products of streaming vectors.  
The instruction set supports hardware-managed, in-memory FIFOs that use memory regions as circular buffers.  The core has special hardware registers to manage the state (head and tail location, for example) of each FIFO.  These are the boxes labeled \texttt{term[0]} to \texttt{term[4]} in the figure and the objects of \texttt{fifo} type in the listing.

Each core exchanges its local iterate vector v with its four neighbors.  The broadcast is performed using a single communication channel that fans out to its four neighbors. But the core receives vector segments on four distinct channels, one corresponding to each of its four neighbors, each consumed by a different background thread handling the four incoming streams.

\begin{figure}[hbtb!]
    \centerline{\includegraphics[width=.72\linewidth]{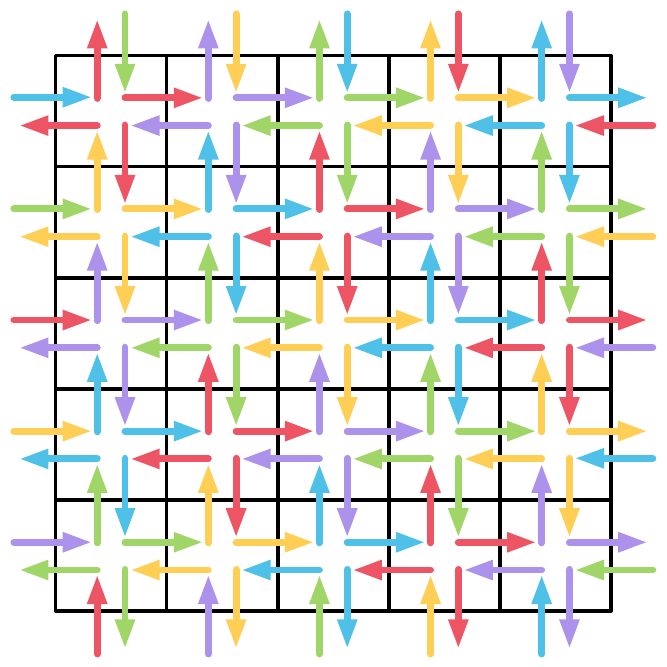}}%
    \caption{Tessellation routing pattern for SpMV: a single core pushes its content into adjacent cores' fabric router using a single communication channel. Messages from the four neighbors arrive on four distinct channels and are processed by corresponding threads. This allows us to achieve high fabric utilization due to the fact that we can send the data in 4 directions in a single cycle. WSE allows the fabric to be dynamically reconfigured. Such adaptive topology plays a significant role in offloading routing logic from cores, which can be used primarily for computation.}
    \label{fig:bcast2}%
\end{figure}

We allocate channel numbers to make all five of these channels different at every tile.  The assignment we used is shown with color as a representative of channel in Figure~\ref{fig:bcast2}.  Note that at every tile, the outgoing color (in four directions) differs from each of the four incoming colors.

Tensor descriptors are used in code to specify the geometry of data and the stepping though it during execution of vector instructions. In Figure~\ref{fig:spmv_flow} these descriptors are annotations, such as \texttt{xp\_rx} (meaning a descriptor of a vector received from the neighbor in the $+x$ direction) next to the arrows between blocks. Some, like \texttt{xp\_rx}, are fabric tensor descriptors that indicate the channel used to communicate this tensor as well as a vector length or tensor shape; they are declared with the fabric keyword in the code.  A memory tensor descriptor such as (\texttt{xp\_a} in the figure and declared with the tensor keyword) can point to a memory address and indicate a length of $Z$ and unit stride for tensors that reside in memory.  

So how does it work?  There is one outgoing stream (the arrow coming out of \texttt{v}, into the router, the dashed box above), and five incoming streams: \texttt{xm\_rx}, \texttt{xp\_rx}, \texttt{ym\_rx}, \texttt{yp\_rx}, and \texttt{c\_rx}, which is the \texttt{v} data looped back. (We loop back the outgoing local data and route it in for processing the $z$ dimension, as this saves memory bandwidth, saving time.)  The assignment statement \texttt{c\_tx[] = v1[]} launches a thread because in the declaration of the fabric tensor \texttt{c\_tx}, there is a thread resource assigned (\texttt{.thr = 5}).

The job is started by a task, called spmv, that first sends the local iterate to the fabric with the assignment mentioned above.
It then initializes the result as the product of the \texttt{zm\_a} memory vector and a shift-by-one, \texttt{v0}, of the in-memory local iterate.   This is done by a tensor multiply instruction (denoted spmvtask in the white box on top) specified by the statement \texttt{zm\_acc[] = v0[] * zm\_a[]}.  Since none of the tensor descriptors has a thread assignment, this runs to completion before spmv proceeds.

The task, spmv, then launches the six background threads (in white boxes, denoted \texttt{thread\_0} through \texttt{thread\_5}) that do elementwise vector multiplication as single SIMD instructions.  These threads pull inputs from the fabric; two of them, 4 and 5, pull the looped-back local data.  They send results to FIFOs (except for thread~5).   

After the main task has launched these background threads, it may safely terminate. The threads it launched continue to multiply and push elements of product vectors into FIFOs.  

Thread 5 simply adds its input to the result vector.  It doesn't need to multiply, because it is working on the main diagonal of the matrix, which is all ones.

The Cerebras hardware FIFOs have a distinctive feature.  They are able to activate tasks, in this case \texttt{sumtask}, whenever they aren't empty.  This they do.   The hardware task scheduler runs \texttt{sumtask} (while the threads continue to fill the FIFOs).  It executes five vector add instructions (as shown in the white box) that each has one FIFO input.  Each add pulls as much data as it can from its input FIFO, finishing when empty. Their destination tensor descriptors track their progress, so that they add once to each of the result tensor.  So \texttt{sumtask} is invoked repeatedly, pulling from the FIFOs, allowing the upstream threads to continue to push data into them.

The six add operations that increment elements of the result run in a temporally interleaved way.  There is no race condition. The additions occur in a nondeterministic order. But there is no danger of a data race, and no locks need be acquired, as the hardware handles the interleaving, working on only one thread at a time.  

In order to return control to the invoking task (called \texttt{bicg} here in the code), we need to ascertain that all five adds are done.   This is done with a small tree of two-way barriers implemented by unblock and activate actions.


\subsubsection{SpMV (2D)}
We sketch an implementation of SpMV ($u = Av$ as above) for a 9-point stencil in 2D. 

For the 2D problem we map a rectangular region of the mesh of $v$ to each core, and store all elements of the corresponding columns of $A$.
After multiplication of the local $v$ with the local $A$ we have generated products in an output halo that must be sent to neighboring tiles.
We accomplish this output-halo exchange with sends of fabric tensors in threads that arrive and feed data into addition threads.
We complete a round of send and add in one direction, then a round for the other direction, and in this way avoid communication along diagonals of the tile grid.


The efficiency of this approach is approximately the same as for the 3D mapping discussed above. On one hand, because all 9 multiplies and adds for a given element of the vector $x$ are performed on the same core, we are able to use the fused multiply-accumulate instruction. The 18 flops performed take place in three machine cycles which increases utilization over the 3D mapping which performed only adds or only multiplies on any given cycle. However, the summation work for the halos are redundant operations that offset this efficiency. Furthermore, although we perform multiplication along the main diagonal, we should not receive performance credit for this operation because most problems will precondition the main diagonal to unity.

The local memory in each core is sufficient to store a matrix, halo, and vector (as well as all terms needed for BiCG) to hold a sub-block up-to 38x38 in size, corresponding to geometries of 22800x22800. Efficiency remains high for smaller problems. When a core holds only an 8x8 region in local memory (4800x4800 meshpoints), the overhead remains less than 20\%.


\subsubsection{AllReduce}

BiCGStab requires inner products of vectors distributed across the whole fabric. This requires an AllReduce operation: the results of local dot products need to be summed across all cores and then the result broadcast back to all cores. Because we did not use a communication-hiding variant of BiCGStab, this collective operation is blocking, so we minimized latency.

\begin{figure}[hbtb!]
    \centerline{\includegraphics[width=.95\linewidth]{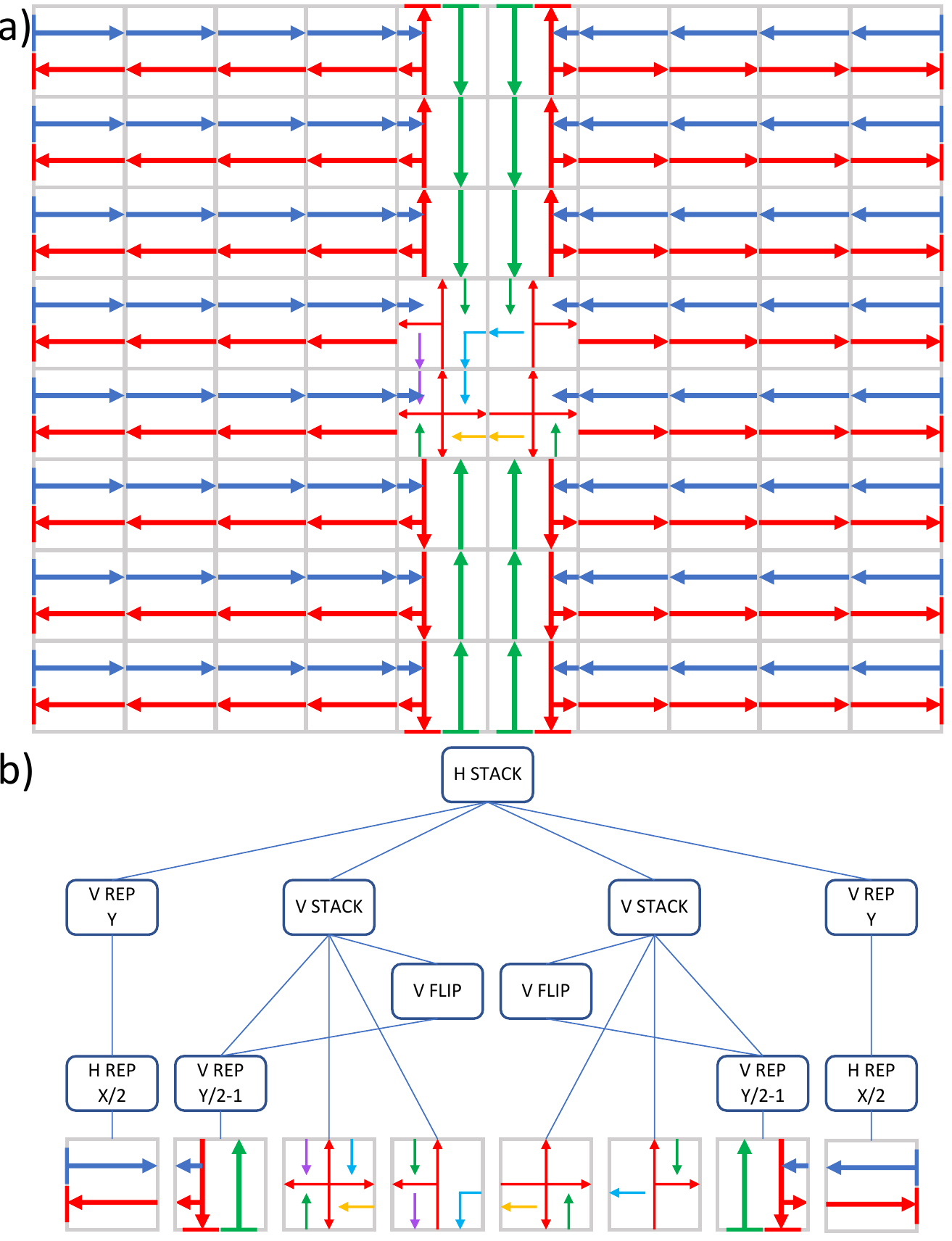}}
    \caption{
(a) The AllReduce operation routing rules for X=8, Y=8. Blue: Horizontal Reduction. Green: Vertical Reduction. Yellow/Cyan/Purple: 4:1 Reduction to a single core. Red: Broadcast result to entire array. 
(b) Route construction. 
   The code builds a DAG of geometry operations (rotation, mirror image flip, and horizontal/vertical stacking) whose leaves are single-tile router configurations, and the DAG is compiled into the fabric routing tables.
    }
    
    \label{fig:allreduce}
\end{figure}

The routing configuration for this task is illustrated in Figure~\ref{fig:allreduce}. 
The reduction is performed in parallel along fabric rows, then along two central columns.  When the reduction starts, each core sends its value toward the center of its row.  On each successive cycle the two central cores of that row receive a datum and accumulate it into a local sum.  We use two cores in the center, each receiving input from one direction at the rate of one datum per cycle.   Similarly, the partial sums are reduced along two columns towards the central four cores  that finally reduce their content to a single core.  The reason for the use of pairs of cores is core-to-fabric injection/extraction bandwidth: a core can add two 32-bit quantities per cycle but can receive only one from the fabric.  The broadcast is done in reverse, sending the result along two central columns and then across all rows.  The single cycle-per-hop latency of the interconnect allows us to implement the AllReduce operation in a cycle count only about 10\% greater than the diameter of the system.


To control the growth of roundoff error, we use a hardware inner product instruction that employs mixed 16-bit multiply/32-bit add precision, and we do the AllReduce at 32-bit precision.

\begin{table}[hbt!]\centering
{\renewcommand{\arraystretch}{1.0}
\begin{tabular}{l|cc|ccc} 

Operation & \multicolumn{2}{c|}{Single precision} & \multicolumn{3}{c}{Half/single mixed} \\
\cline{2-6}
 (x Count)      &              SP + & SP $\times$ & HP + & HP $\times$ & SP + \\
\hline
Matvec (x2)  & 12  & 12 & 12 & 12 & 0 \\ 
Dot  (x4)  & 4  & 4 & 0 & 4 & 4\\ 
AXPY (x6)  & 6  & 6 & 6 & 6 & 0 \\ 
\hline
Total & 22 & 22 & 18 & 22 & 4 \\

\end{tabular}}
\vspace{4pt}
\caption{Operations per meshpoint per iteration}
\label{table:ops}
\end{table}

%

\subsubsection{AXPY}
These operate on core-local fp16 data and use the four-way SIMD capability.

\section{Measured Results}
We implemented BiCGStab for a 
$600 \times 595 \times 1536$ mesh on a CS-1 machine available to us for experiments that has a $602 \times 595$ compute fabric.
We measured the wall clock time between consecutive BiCGStab iterations; the mean differences across $171$ iterations was $28.1$ microseconds.  The standard deviation is about 0.2\% of the mean.

Per Table I, there are $44$ operations per meshpoint per BiCGStab iteration.  Thus, we perform $44 \times 600 \times 595 \times 1536$ operations per iteration floating-point, which works out to an achieved performance of 0.86 PFLOPS.   As pointed out in Table II, we are using 16-bit floating point for $40$ operations per iteration per meshpoint, and 32-bit floating point for the remaining $4$ operations.

\subsection{A strong scaling comparison}
\begin{figure}[hbt!]\centering

\begin{minipage}[b]{0.4\textwidth}
  \includegraphics[width=\textwidth]{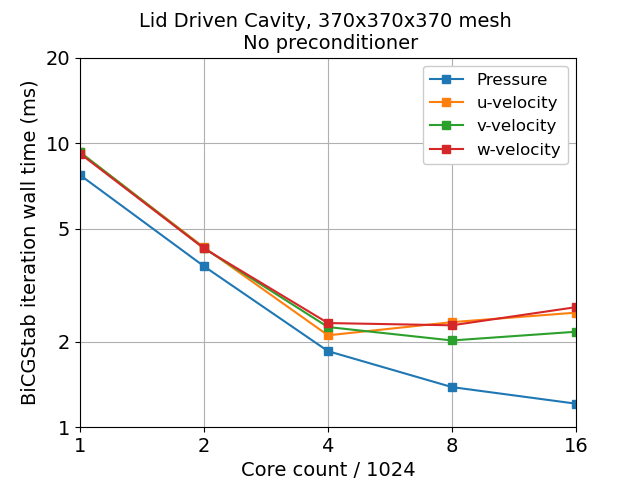}
  \caption{Scaling of solve time on a cluster, $370^3$ mesh}
   \label{fig:jouleLDC370}
\end{minipage}
\begin{minipage}[b]{0.4\textwidth}
   \includegraphics[width=\textwidth]{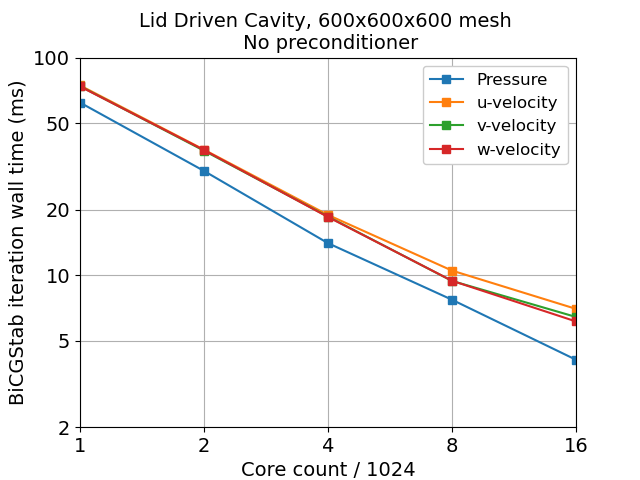}
   \caption{Scaling of solve time on a cluster, $600^3$ mesh}
   \label{fig:jouleLDC600}
\end{minipage}

\end{figure}

To make a comparison to CPU-based clusters, we solved a comparable problem, the BiCGstab solution of a nonsymmetric linear system arising from a 7-point stencil finite volume approximation; this was done within the NETL MFIX code while computing a lid-driven cavity flow. We have data for two problem sizes, both $370^3$ and  $600^3$. For the larger mesh, the performance for systems with numbers of cores ranging from $1024$ to $16384$ is shown in Figure~\ref{fig:jouleLDC600}.  These are 64-bit floating point results obtained on Joule 2.0, the NETL supercomputer, which is based on HPE ProLiant servers, Intel Xeon Gold 6148, 20-core, 2.4GHz processors, using the Intel Omni-Path interconnect.  (In a Navier-Stokes solver like MFIX, four linear systems are solved at every time step, one for each of the solution variables, three velocity components $u, v, w$ and pressure $p$.)  There are some differences between the behavior for the pressure and the momentum equations, but these differences do not change the main points.


The failure to scale beyond 8K cores on the smaller mesh highlights the difficulty of achieving strong scaling on a cluster.

For the larger mesh, time per BiCGstab iteration on Joule ranges from 75 ms on 1024 cores, and scales down to about 6 ms on 16K cores.   This is about 214 times more than the $28.1$ microseconds per iteration that we measured on the CS-1, on a problem with more than twice as many meshpoints.   (On the other hand, the arithmetic is four times wider on Joule.)

It is interesting to try to understand why this striking difference arises.   A first explanation is that there are a lot more cores on the CS-1.  The cores are not equal in their performance, however, and the peak performance of 16K cores on Joule at 64-bit is about 40 percent of the CS-1 peak at 16-bit.   This explains some of the difference, but there is still a large performance gap per core.

The difference in memory technology is clearly important.  A single 20-core Xeon 6148 socket has  27.5 MB of last-level (L3) cache, and at 16K cores the aggregate cache is 22.5 GB, substantially more than the memory of the CS-1.  But the Xeon caches seem to be less effective at deriving performance from the available SRAM.   We suspect that this may be because the L3 cache is shared by the 20 cores on the socket, and does not have enough port bandwidth to satisfy all of their needs when, as in this case, there will be little temporal data reuse in the L1 and L2 caches.   The CS-1 SRAM is not shared, and each core has access to an array of small SRAM banks that deliver 16 bytes (read) and 8 bytes (write) per clock to the single core, which is enough to support the full compute rate for an operation like an AXPY that streams two vectors from memory and streams the result vector back.





\section{CFD on the CS-1}

Real world applications exhibit complexity not found in simple incompressible Euler flow in rectangular geometry; they feature complex geometries with heat, mass, compressibility, stretched meshes, and even structure interactions.  These all complicate the steps to form the matrices of the linear systems. They account for 30 to 50 percent of the operation count. And they add to the memory requirement---memory will limit the maximum problem size that can be solved on CS-1.  Thus, we have to look at what happens outside the linear solver.  We analyze next the work and memory required in a realistic case.

\begin{algorithm}
\caption{SIMPLE in MFIX}
\label{alg:simple}
\begin{algorithmic}[1]
    \State Initialization (calculate shear and time dependant source)
    \For{i = 0,1,2, ...}
    	\For{ii = u,v,w}
    	\State Form Momentum
    	\State BiCGStab Solve
    \EndFor
    \State Form Continuity
    \State BiCGStab Solve Continuity
    \State Field Update (u, v, w, p)
    \State Calculate Residual
    \EndFor
\end{algorithmic}
\end{algorithm}

As a case study, we examine the Multiphase Flow with Interphase eXchanges (MFIX) CFD code from the National Energy Technology Laboratory.  In particular, we examine the newest variation on the code MFIX-TF, which is a fully vectorized formulation.  MFIX is a general purpose, Cartesian mesh, multi-phase CFD code.  The code solves the fully compressible Navier-Stokes equations using an adapted version of the Semi-Implicit Method for Pressure Linked Equations (SIMPLE).  In order to show that a significant problem can be solved, we discuss a single phase, compressible, viscous fluid problem without energy and species equations.  It is straightforward to extrapolate the allowable size and arithmetic intensity at any level of complexity following the methodology outlined below.

\subsection{Performance and accuracy of CFD on the CS-1}
The operations needed to construct the coefficient matrix and source vector depend on the discretization scheme.  First order upwinding is the most common scheme and was used to determine operation types and counts.  
The necessary operations can be grouped into vector merge operations, floating point (FLOP) operations (multiply, add, subtract), square root, divide, and neighbor transport operations.  The cycle counts for each operation have been estimated and the operations counted for all steps outside of the linear solver in the SIMPLE algorithm per $Z$ meshpoint.  The residual calculations were ignored because they involve dot products and a few scalar calculations.  The analysis revealed that they could be overlapped with other computations.

\begin{table}[hbt!]\centering
\begin{tabular}{ |l|c c c c c c| } 
\hline
SIMPLE Step & Merge & FLOP & $\sqrt{}$ & $\div$ & $\mathbf{x}^\intercal$ & Total\\  
\hline
Initialization & 2 - 9 & 35-47 &0 &0 & 8& 45-64\\ 
Momentum & 25-153 & 18-25& 13 & 15-16 & 6&79-213\\
Continuity & 8-45 & 13-18& 0 & 15-16&2&37-81\\
Field Update& 0 & 3-5 & 0 & 0&1&4-6\\
\hline
\end{tabular}
\vspace{4pt}
\caption{Cycles per meshpoint for SIMPLE, excluding the solver.  
}
\label{table:nonlsCost}
\end{table}


For typical flow problems, the number of simple iterations ranges from 5-20 per time step, the linear solver is limited to 5 iterations for transport equations and 20 for continuity equation.  Based on the performance estimates, the wall time per time step was estimated to be roughly two microseconds per $Z$ meshpoint.  Assuming a problem size of 600x600x600 and 15 simple iterations per time step, and we expect to achieve between 80 and 125 timesteps per second.  This places the likely performance of CS-1 above 200 times faster than for MFiX runs on a 16,384-core partition of the NETL Joule cluster.   Furthermore, our experiments indicate that little more performance can be gained by scaling the cluster with fixed mesh size.

\subsection{Accuracy achieved with mixed precision arithmetic}
There is already extensive work on the use of lower precision in linear algebra and other areas.
Ongoing research has explored how full precision can be maintained while doing most work in lower precisions~\cite{carson, precimonious, 10.1007/978-3-319-93698-7_45}.  For linear systems, variants of iterative refinement have been successful. 

\begin{figure}[hbt!]\centering
  \includegraphics[width=.93\linewidth]{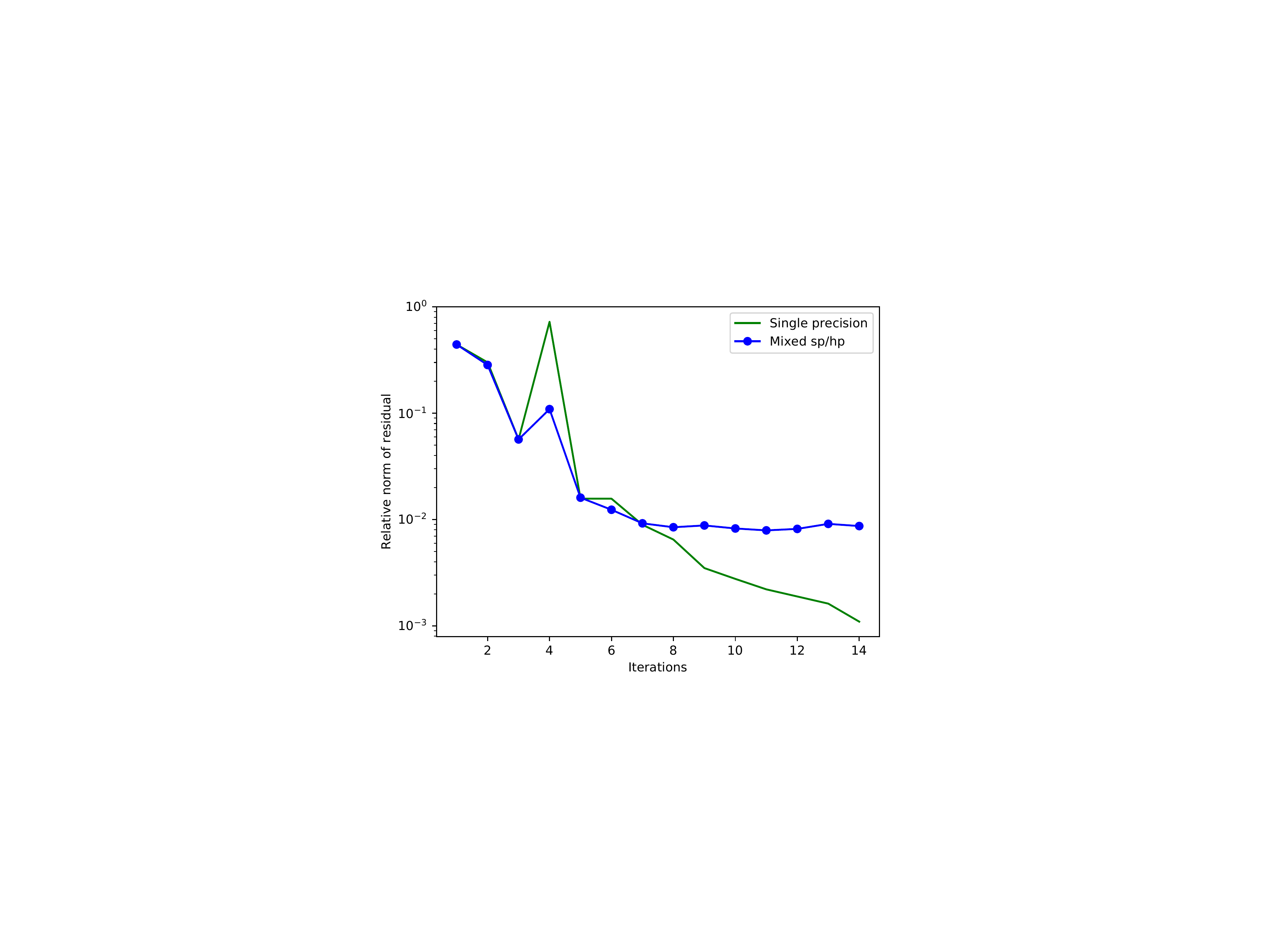}
  \caption{Normwise relative residual in mixed and 32-bit arithmetic.   Mixed precision plateaus as expected near its machine precision.}
  \label{fig:residual_80}
\end{figure}

We have assessed the quality of solution we obtain using mixed 16 and 32-bit precision.  
We took a linear system from the timestep discretization (in the NETL code MFIX) of the momentum equation for a velocity component on a $100 \times 400 \times 100$ mesh.  The measured normwise relative residuals with mixed and 32-bit are shown in Figure~\ref{fig:residual_80}. Up to iteration 7 the mixed precision implementation tracks the 32-bit, but then fails to reduce the residual further.
With this precision, machine precision is about $10^{-3}$.  We have observed this accuracy for very well conditioned systems.   Here, the growth of rounding errors during the iterative solve explains the loss of an additional factor of $10$, leading to a plateau at a relative residual of $10^{-2}$.  We have not yet tested whether this accuracy is acceptable in the MFIX code.

We expect that for some realistic situations, mixed precision solvers are usable as is; in others they may need to be coupled with a correction scheme such as an iterative refinement or an outer iteration that solves a nonlinear system; and in other situations one may need to use higher precision arithmetic.  There is interest in finding the limits.  In a 2018 Gordon Bell Finalist the authors demonstrate a nonlinear finite-element solver in which linear systems are solved using mixed-precision implementations based on fp16 arithmetic~\cite{10.1109/SC.2018.00052}.  
A large-scale example is NASA's Fun3d code.  NASA is using Fun3d to simulate the powered supersonic entry of manned space vehicles in the Martian atmosphere.  To reduce runtimes, NASA has implemented cast in place kernels to convert fp32 values to fp16 so that they can run the Red-Black Gauss-Seidel linear solver in fp16 and transition to using Tensor Cores.  They reported no ill effects on convergence or stability\cite{nasa16}.

\section{Related Work}

In contrast to this work, there has been extensive work on specialized machine aimed at particular use cases.   Examples include Anton~\cite{anton} and the MDGRAPE systems~\cite{mdgrape4}.   This work stands in stark contrast to ours.   We explain that a new approach at the hardware layer provides a breakthrough in memory bandwidth that make a fully general purpose system, the CS-1, capable across the spectrum of arithmetic intensity.   The efforts we cite provide specialized machines that are ordinary at the hardware level and that draw their advantages from architectural specialization, and that are aimed at molecular dynamics, which lives at the high end of the intensity spectrum.

The issue of memory and fabric communication in PDE and Krylov subspace solvers remains unresolved for conventional systems.  Indeed it has spawned research on communication avoiding methods~\cite{hoemmen, carsonthesis}.   These can produce incremental improvements, but memory bandwidth remains a performance limiting issue.


\section{Discussion and Open Questions}
\subsection{Applications}
We have shown here that the approach of keeping all the processing, data, and communication on one silicon wafer can, when a problem fits on the wafer, eliminate memory bandwidth and greatly reduce communication as performance limits, yielding dramatic speedup and allowing accurate real-time simulation.  Before now, practitioners have had to choose between very fast reduced order/fidelity models which often lack the resolution needed to be really useful and high-resolution simulations which lack the speed to be practical. 

To our knowledge, this is the first ever system capable of faster-than real-time simulation of millions of cells in realistic fluid-dynamics models. In real-time applications, memory footprints are normally tractable in order to achieve the mandatory real-time performance.  To take one example, it is quite difficult and potentially dangerous to land a helicopter on the windy flight deck of an aircraft carrier, due to the complexity and speed of the air flow.  In his 2017 PhD thesis~\cite{oruc}, Oruc noted that ``Especially, when
the aircraft flies in the near vicinity of the superstructure of the ship, the pilot
workload increases seriously.''  He examines the potential for semi-automatic flight control with in the loop CFD.   He found that modest meshes of in the neighborhood of one million cells can provide adequate accuracy, but that the necessary real-time performance is hard to achieve on a cluster of multicore CPU systems.  We give other examples of compact applications in the section below.

Cheap simulations mean that they could be treated as an input to a neural network residing in CS-1 as well and train a model approximating them. It has been shown that combining machine learning with traditional simulation accelerates scientific experiments\cite{Huntingford_2019}. Another implication is that we could simulate an event on demand instead of storing previous experiments. This solution makes interactive fluid dynamics a reality.

In NETL's application domain, real-time simulation can enable physics-based online equipment monitoring, cyber-physical security, equipment failure prediction, dynamic baseload power following, fixed asset turn down, and renewable integration. 
It may become less expensive to rerun simulations than to save and retrieve information from storage.  
And, since high efficiency operation of energy systems often depends on operating near failure points, the advent of faster than real time, high resolution simulations may make it safe to achieve unprecedented levels of efficiency. 

\subsection{Memory capacity}
There are compelling HPC use cases for the \mbox{CS-1}, notwithstanding its modest memory capacity.
In the helicopter example of the last section and in three examples below, extraordinary performance in a limited memory footprint is of great value. In these examples a great many spatially compact problems must be solved, for purposes that span uncertainly quantification, to long time duration simulations involving perhaps millions of timesteps, to real-time deployments with in-the-loop models.  None of these use cases can be addressed using systems based on low-performance, low cost-per-byte memory.

The first use case is in automated exploration of design spaces.  
In the work of Madsen et al~\cite{madsen}, 
the first attempt at 3D simulation-based shape optimization of whole wind turbine rotors, the wind turbine scale was 10MW.  Richardson extrapolation showed that the number of cells needed is in the 14-50M range to get estimates within 0.5\% to 2\%.  In the optimization setting, simulations have to be done sequentially, as the optimizer depends on gradients around simulation runs.  There can be hundreds to thousands of simulations to complete a shape optimization.  This type and scale of problem is also common in aircraft and automotive design optimization.  Today these optimizations are not routinely done due to the compute times.

Second, as computational models play a growing role in industrial and societal decision-making, the confidence worthiness of these models has to be quantified.  Uncertainty quantification (UQ) relies on the assessment of the effect of model and parameter uncertainty, usually requiring hundreds or thousands of simulations to sweep relevant ranges of parameters. For example, to predict the performance uncertainty of a 1MW pilot scale carbon capture system providing 90\% capture efficiency with 95\% confidence~\cite{xu}, a total of 1,505 simulations were run, each needing about 600 seconds of simulation to reach steady state. This kind of simulation campaign is currently achievable only by simplifying the geometry (two-dimensional geometry), and using filter models to account for the coarse meshes. 

Finally, consider a study by Jasak et al~\cite{jasak} of commercial ship modeling in which speed rather than capacity is the issue.  The solution time on a moderately sized computer system typical in engineering was as much 83 hours for a test case involving 11.7 million finite volume cells.  Such a turnaround time is clearly a burden in design optimization and design space exploration.

Our goal is not to tout the CS-1 specifically; rather we want to illuminate the benefits of the wafer-scale approach. Recall that the CS-1 is the first of its kind; we expect that memory limits will recede over time with ongoing technological and engineering developments.  A technology shrink from the 16 nm to 7 nm technology node will provide about 40 GB of SRAM on the wafer and further increases (to 50 GB at 5 nm) will follow.  These changes will help, and while they don't produce petabyte capacities,
there are new directions that the hardware can take, involving systems that go beyond a single wafer and a single wafer type, that can add much more.  Solutions involving the clustering, with sufficient bandwidth, of several wafer-scale systems is certainly a possibility.

\section*{Acknowledgement}
The authors would like to thank Natalia Vassilieva for initiating the collaboration between Cerebras Systems and NETL and for her subsequent help with the project. 

\bibliographystyle{IEEEtran}
\bibliography{main}




\end{document}